\documentclass[sigconf]{acmart}
\acmConference[CAIN 2024]{3rd International Conference on AI Engineering — Software Engineering for AI}{April 2024}{Lisbon, Portugal}%% Fonts used in the template cannot be substituted; margin 
\usepackage{listings}
\usepackage{xcolor}

\AtBeginDocument{%
  \providecommand\BibTeX{{%
    \normalfont B\kern-0.5em{\scshape i\kern-0.25em b}\kern-0.8em\TeX}}}

\setcopyright{acmcopyright}
\copyrightyear{2018}
\acmYear{2018}
\acmDOI{XXXXXXX.XXXXXXX}

\acmConference[Conference acronym 'XX]{Make sure to enter the correct
  conference title from your rights confirmation emai}{June 03--05,
  2018}{Woodstock, NY}
\acmBooktitle{Woodstock '18: ACM Symposium on Neural Gaze Detection,
 June 03--05, 2018, Woodstock, NY} 
\acmPrice{15.00}
\acmISBN{978-1-4503-XXXX-X/18/06}

\begin{document}

\title{ML-On-Rails: Safeguarding Machine Learning Models in Software Systems – A Case Study}


% \author{Hala Abdelkader, Mohamed Abdelrazek, Scott Barnett, Jean-Guy Schneider, Priya Rani, Rajesh Vasa}
% \email{habdelkader@deakin.edu.au, mohamed.abdelrazek@deakin.edu.au, scott.barnett@deakin.edu.au, jean-guy.schneider@monash.edu, priya.rani@rmit.edu.au, rajesh.vasa@deakin.edu.au}
% \affiliation{\institution{Deakin University \country{Australia}}  \institution{Monash University \country{Australia}} \institution{RMIT University \country{Australia}}}

\author{Hala Abdelkader}
\email{habdelkader@deakin.edu.au}
\affiliation{
  \institution{Deakin  University}
  \country{Australia}
}
\author{Mohamed Abdelrazek}
\email{mohamed.abdelrazek@deakin.edu.au}
\affiliation{
  \institution{Deakin  University}
  \country{Australia}
}
\author{Scott Barnett}
\email{scott.barnett@deakin.edu.au}
\affiliation{
  \institution{Deakin  University}
  \country{Australia}
}
\author{Jean-Guy Schneider}
\email{jean-guy.schneider@monash.edu}
\affiliation{
  \institution{Monash University}
  \country{Australia} }
\author{Priya Rani}
\email{priya.rani@rmit.edu.au}
\affiliation{
  \institution{RMIT University}
  \country{Australia}
  }
\author{Rajesh Vasa}
\email{rajesh.vasa@deakin.edu.au}
\affiliation{
  \institution{Deakin  University}
  \country{Australia}
}
% \affiliation{%

% \affiliation{%

% \affiliation{%

% \affiliation{%

% \affiliation{%

% \affiliation{%

% \affiliation{%

\renewcommand{\shortauthors}{Trovato and Tobin, et al.}

\begin{abstract}
  A clear and well-documented \LaTeX\ document is presented as an
  article formatted for publication by ACM in a conference proceedings
  or journal publication. Based on the ``acmart'' document class, this
  article presents and explains many of the common variations, as well
  as many of the formatting elements an author may use in the
  preparation of the documentation of their work.
\end{abstract}






\begin{abstract}
Machine learning (ML), especially with the emergence of large language models (LLMs), has significantly transformed various industries. However, the transition from ML model prototyping to production use within software systems presents several challenges. These challenges primarily revolve around ensuring safety, security, and transparency, subsequently influencing the overall robustness and trustworthiness of ML models. In this paper, we introduce ML-On-Rails, a protocol designed to safeguard ML models, establish a well-defined endpoint interface for different ML tasks, and clear communication between ML providers and ML consumers (software engineers). ML-On-Rails enhances the robustness of ML models via incorporating detection capabilities to identify unique challenges specific to production ML. We evaluated the ML-On-Rails protocol through a real-world case study of the MoveReminder application. Through this evaluation, we emphasize the importance of safeguarding ML models in production.

% Machine learning (ML), especially with the advent of large language models (LLMs), has significantly transformed various industries. However, transitioning from ML model prototyping to production use in software systems poses multiple challenges around the robustness of the ML models and the lack of standardization in ML model management tools and processes. In this paper, we introduce ML-On-Rails: ML safeguarding protocol that establishes well-defined endpoint interface for different ML tasks, and clear messaging between ML developers and ML users (software engineers). robustness handling capabilities, an with detection capabilities to identify unique challenges specific to production ML. We evaluated ML-On-Rails protocol using a real-world case study MoveReminder application as a case study. Through this evaluation, we emphasize the importance of safeguarding ML models in production.  \todo{add few more details about your evaluations, and KEY FINDINGS} 
% \todo{link and mention the tool here}

\end{abstract}
\maketitle

\section{Introduction}
Machine learning (ML) models have evolved into key components in various real-world applications, each with its own level of impact and complexity~\cite{goodfellow2016}. These technologies are extensively applied in human-centered systems, such as healthcare~\cite{balas2023conversational, feng2023large, shyr2023identifying, zhou2023skingpt, abobakr2018rgb}, autonomous driving~\cite{ bojarski2016, chen2015deepdriving, eykholt2018robust, tian2018}, and human pose estimation~\cite{ abobakr2019}. However, ML models face challenges that impact the overall software system robustness~\cite{amershi2019software, kumeno2019sofware, schelter2018challenges, guo2019, gopinath2019}. 

Robustness concerns towards ML-enabled systems have been categorised into three main aspects; transparency, safety, and security~\cite{arnold2019factsheets, hendrycks2021unsolved}. Safety includes elements such as explainability, fairness, and robustness to dataset shift, while security includes defence against adversarial attacks~\cite{arnold2019factsheets, hendrycks2021unsolved}. Transparency of ML models has also been a focus area to avoid the misuse of ML models and preserve their trustworthiness~\cite{mitchell2019, arnold2019factsheets, hendrycks2021unsolved, raji2019ml}.

Further to the above concerns is the silent failures of ML models. It is an inherent behaviour where ML models generate outputs without signaling expected errors or warnings, regardless of input validity~\cite{rabanser2019}. This lack of explicit alerts poses a significant challenge when deploying ML models in production systems as it complicates the identification and resolution of performance issues. This silent failure, if left unaddressed, can lead to profound consequences, undermining the robustness and efficacy of the entire system.

The recent emergence of large language models (LLMs) and their expanded application domains, further robustness challenges have been introduced. LLMs take a prompt as input and generate a set of tokens through sampling from a probability distribution. LLMs can easily be manipulated to generate off-topic responses~\cite{pang2023leveraging}. Additionally, they tend to generate responses that are factually incorrect or entirely fabricated, referred to as hallucinations~\cite{manakul2023selfcheckgpt, peng2023check, azaria2023internal}. Furthermore, LLMs are vulnerable to prompt injection attacks, where malicious actors manipulate inputs to deceive the model into producing harmful outputs~\cite{kang2023exploiting, wei2023jailbroken, zou2023universal}.

While traditional software engineering has established tools and methodologies for maintaining system robustness, the fundamental differences between ML and traditional software components make it challenging to apply these practices to ML models~\cite{amershi2019software, kumeno2019sofware}. Unlike traditional software, ML components rely on data-driven learning techniques that use training data to approximate a mapping function or logic to use for inference~\cite{kumeno2019sofware}. There is ambiguity surrounding the learned logic making it difficult to adopt traditional software engineering tools for ML components. 

To ensure the robustness of ML-enabled systems, it is crucial to safeguard integrated ML models and maintain a continuous monitoring and evaluation of the deployment environment~\cite{schelter2018challenges, baier2019challenges}. In this paper, we propose ML-On-Rails, a protocol designed to safeguard ML models. It features detection of production ML problems, communicates performance issues, and enforces specifications of ML model. We evaluate the proposed protocol through a case study of a MoveReminder application, a mobile health solution for individuals diagnosed with type-2 diabetes. The application objective is to motivate a more active lifestyle through personalized activity recommendations. It incorporates two ML models: an activity recognition model and the GPT-3.5 LLM. The activity recognition model uses sensory data for classification of the user's physical activity. Additionally, the application gathers user preferences and other attributes, including weather, location, date, and time. This data is then fed to the LLM to generate activity recommendations.

Research questions for this work include:

\begin{itemize}
    \item \textbf{RQ1} What are the key components necessary to safeguard ML models in production environments?
    \item \textbf{RQ2} What protocols we need to effectively communicate between the ML engineering side, and the software engineering side?
\end{itemize}

Contributions of this work are summarized as follows.
\smallskip

\begin{itemize}
    \item We introduce the ML protocol, a novel framework designed to enforce ML model specifications and enhance elements of robustness in ML-based software systems.
    \item We introduce a set of key ML safeguard requirements and existing methods to achieve these requirements.
    \item We demonstrate the ML protocol through a real-life case study, highlighting its applicability and benefits in real-world scenarios.
\end{itemize}

The rest of this paper is organized as follows: in Section~\ref{sec:related_work}, we provide an overview of research related to the robustness of ML. 
In Section~\ref{sec:ml-protocol}, we introduce the ML protocol and the case study is discussed in Section~\ref{sec:study-design}. Finally, The paper concludes with a summary of the key observations in Section~\ref{sec:conclusion}.

\section{Literature review}
\label{sec:related_work}

% What is NeMo Guardrails
The concept of establishing a standardized protocol for ML models has not been extensively explored in the literature. The most relevant work is NVIDIA's NeMo Guardrails for generative models applications. NeMo Guardrails is an open-source toolkit designed to simplify the integration of programmable guardrails into LLM-based conversational systems to control the output of LLMs~\cite{rebedea2023nemo} to be trustworthy, safe, and secure.

% Who is responsible for adding the guardrails? 
There are various ways for LLM providers and developers to incorporate guardrails into an LLM-based conversation system. Guardrails can be embedded during LLM training with model alignments techniques including RLHF and RLAIF. Prompt engineering techniques such as chain-of-thought (CoT) can also be used to implement the guardrails. However, these approaches are model specific and resource intensive. In contrast, NeMo Guardrails allows developers to include programmable rails in LLM applications at runtime. These rails are user-defined, operate independently of the underlying LLM, and are easily interpretable~\cite{rebedea2023nemo}. 

% Types of rails 
Developers can implement two main categories of programmable rails: topical rails and execution rails. Topical rails are designed to guide responses for specific topics or implement complex dialogue policies. On the other hand, execution rails are custom actions defined by the application developer to monitor LLMs input and output. Execution rails focus on LLMs safety, for example rails for fact-checking, hallucination, and moderation~\cite{rebedea2023nemo}. 
\smallbreak
% Limitations
NeMo Guardrails aims to create safe and controllable LLM applications using programmable rails. However, it is not meant to be a stand-alone solution, especially for safety-specific rails. The recommendation is to combine programmable and embedded rails to build secure applications. The Guardrails runtime CoT prompting approach introduces extra costs and delays. Currently, it results in about three times the delay and cost compared to a regular message generation without Guardrails. However, it gives the developers different choices to meet their specific needs.
\smallbreak
We will incorporate NeMo Guardrails into the ML-On-Rails protocol to provide a complete solution for ensuring the robustness of ML-enabled software systems. The proposed protocol addresses robustness concerns of traditional ML models and incorporates NeMo Guardrails to support the emerging generative LLM models.

\section{ML-On-Rails}
\label{sec:ml-protocol}
This section presents the ML-On-Rails Protocol, comprising two key components. First, a suite of ML Safeguards is outlined, defining and enforcing robustness attributes. Second, the Model-2-Software communication protocol is detailed, encompassing essential messages for embedding errors related to robustness. Finally, we will explore diverse implementation patterns for ML safeguards.
\begin{figure*}
\includegraphics[width=.75\linewidth]{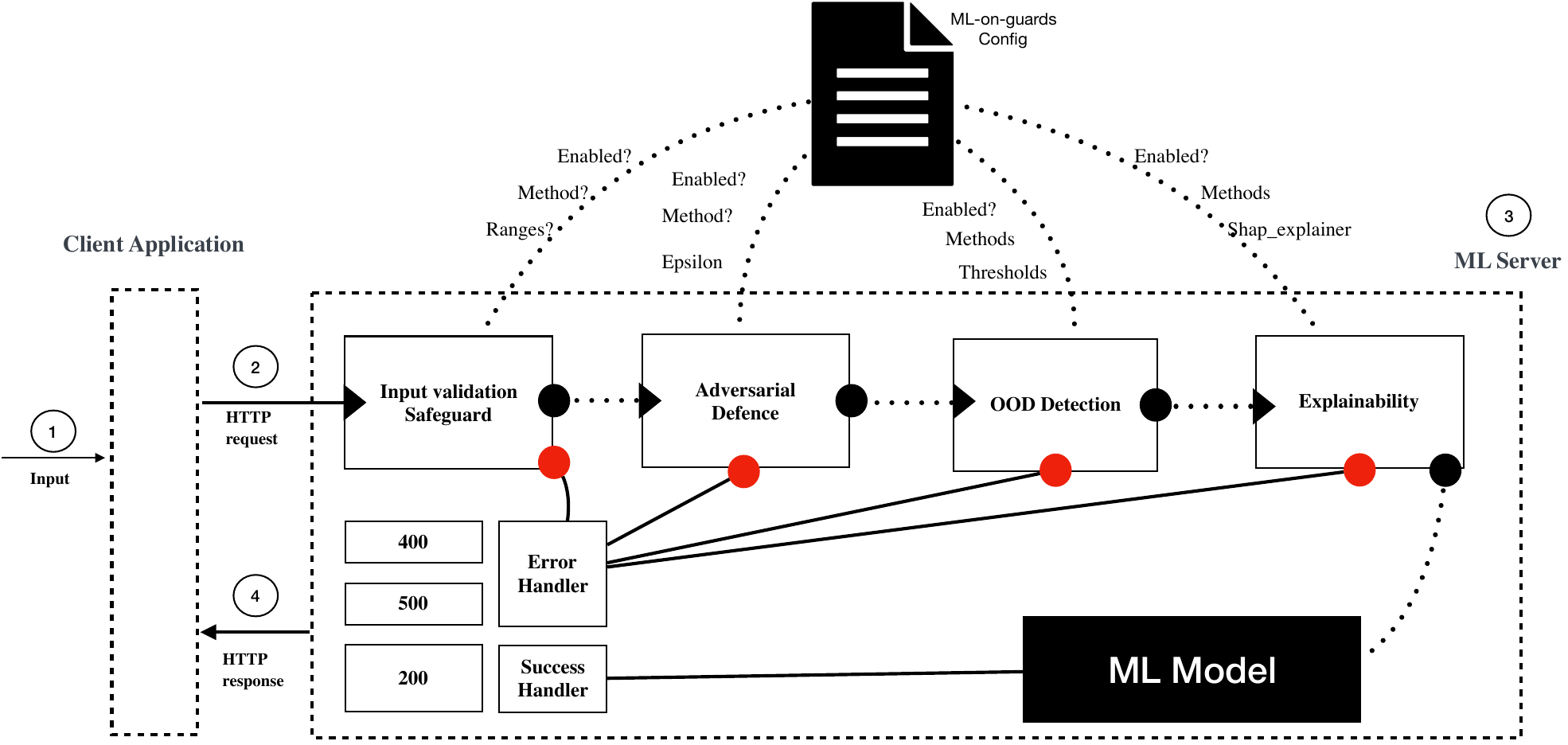}
  \caption{The proposed ML-On-Rails protocol. \textmd{\textit{We propose that ensuring robustness of ML-enabled systems requires safeguards for input validation, security; exemplified by the adversarial defence rail, safety; exemplified by OOD detection rail, and explainability. It is important to highlight that the proposed protocol components is a suggested design, providing developers with the flexibility to enable, configure, disable specific guards, or introduce additional guards. The outlined sequence of guard execution starts upon receiving an HTTP request from a client application. The process involves validation of the input against model requirements. Subsequent checks are conducted for adversarial defence and OOD detection. Once these checks pass, the model inference is performed, followed by the execution of the explainability guard.}}}
   \label{fig:ML_safeguards}
   \vspace{-4mm}
\end{figure*}

\subsection{ML Safeguards}
We translate the robustness concerns of ML models; safety, security and transparency into safeguard components. This process is essential as it lays the foundation for constructing a thorough and efficient framework to systematically address and mitigate potential robustness risks. An overview of the ML-On-Rails Protocol safeguards is illustrated in Fig~\ref{fig:ML_safeguards}.

\noindent\textbf{Adversarial attack detection:} While there has been substantial focus on securing ML-enabled systems in the software industry~\cite{responsible_ai_practices, securing_ai_microsoft, adversarial_ml_ibm}, a recent survey highlights a significant gap. The majority of industry practitioners lack the necessary tools to secure their systems against adversarial attacks in particular, emphasising a clear need for guidance~\cite{kumar2020adversarial}. Concurrently, extensive research has been conducted to enhance the defense mechanisms for ML models against adversarial attacks. To address this disparity, we propose the integration of an adversarial attack safeguard using the adversarial training approach~\cite{goodfellow2014, szegedy2014intriguing, lyu2015unified}. This method involves augmenting the training set with perturbed data. The ML provider can train the ML model on the expanded data to create a robust model or train a separate model for detecting adversarial attacks, or combine both methods. Considering the dependency on training data, adversarial training is employed by ML providers to improve model robustness.

\noindent\textbf{Out of distribution (OOD) detection:}
In production ML-enabled systems, OOD data detection has emerged as a crucial aspect for ensuring the safety of ML models~\cite{yang2021generalized}. OOD data refers to input data that deviates from the training data used to train the model~\cite{yang2021generalized, hendrycks2017}. The presence of such data poses a risk as it can induce ML models to provide inaccurate and overly confident predictions~\cite{liang2018}, leading to substantial consequences in real-world applications, where wrong predictions can have severe implications. Therefore, the proposed ML-On-Rails protocol incorporate OOD detection safeguard to detect OOD inputs that lie outside the training data distribution. There are various methods for OOD detection~\cite{yang2021generalized}, such as the baseline Maximum Softmax Probability(MSP)~\cite{hendrycks2017}, ODIN~\cite{liang2018}, energy-based~\cite{liu2020energy}, DICE~\cite{sun2022dice}, ReAct~\cite{sun2021react} and others.

\noindent\textbf{Model explainability:}
Model explainability contributes to building trust in ML-enabled systems. In production, understanding the rationale behind a model's predictions is crucial for debugging, enhancing model performance, and addressing potential biases. By providing a deeper understanding of model behavior, explainability not only ensures the robustness of ML-enabled systems but also facilitates seamless integration into real-world applications. The explainability safeguard in our protocol uses SHAP (Shapley Additive exPlanations)~\cite{lundberg2017unified}. SHAP focuses on revealing how individual features impact predictions. Derived from cooperative game theory, SHAP Values create a comprehensive framework for assessing feature importance in predictions. 

% In computer vision tasks, input validation is designed to ensure that images meet the criteria for accurate model inference.
\noindent\textbf{Input validation:} Input validation safeguard is a critical component of ML-enabled systems, ensuring accurate data for model inference. In computer vision tasks, validation checks include dimensionality validation, and assessment of quality, contrast, and resolution to filter poor images. In NLP tasks, example input validation steps include; language filtering based on supported language(s) using language identification algorithms~\cite{toftrup2021reproduction, jauhiainen2019automatic, tromp2011graph}, misspelling detection and correction~\cite{hu2020misspelling}, and text length validation.

\noindent\textbf{Generative AI Safeguards:}
For ML-enabled systems incorporating generative AI capabilities and LLMs, the challenges have further dimensions. It is crucial to guarantee safe, secure, and ethical responses from the model. Our proposed protocol integrates the NeMo Guardrails framework to safeguard the LLM components. NeMo Guardrails will address prevalent risks associated with LLMs, such as hallucinations, divergence from the intended application use case, and prompt injection. 
\begin{table*}[htbp]
    \centering
    \begin{tabular}{c p{0.3\linewidth} p{0.5\linewidth}} 
        \hline
        \textbf{HTTP status} & \textbf{Attributes} & \textbf{Possible values} \\
        
        \hline
        200 Ok & message & Success \\
        & status\_code & {HTTPStatus.OK} \\
        & data: class & 
            A List of values for model classes \\
        & data: confidence & A list of values for model confidences \\
        & explainability & Text or image explaining model decision \\
        & source & String to indicate the source of the response. \\
        \hline
        400 Bad Request & message & The server encounters difficulties processing the input for the specified "attribute". The provided data does not meet the expected format or requirements, leading to a processing error on the server side. \\ 
        
        & status\_code & HTTPStatus.BAD\_REQUEST \\

        & error\_code & INVALID\_ATTRIBUTE\_TYPE \\
        
        & type & BadRequest \\
       
       & source &  String to indicate the source of the error. \\
       
       & data: attribute & String to indicate the attribute causing the error. \\
       
       & data: expected\_value & The value expected by the model inference flow. \\

       & data: value & The actual value provided to the model inference flow. \\

       & request-id & String to represent the unique identifier for the request. \\
       & error-id & String to represent the unique identifier for the error. \\
        \hline
        
        500 Internal Error & message &The server detected a problem with the provided input. The provided input distribution does not meet the requirements, leading to a processing error on the server side. \\ 
        
        & status\_code & HTTPStatus.INTERNAL\_SERVER\_ERROR \\

        & error\_code & OUT\_OF\_DISTRIBUTION, ADVERSARIAL\_ATTACK\_DETECTED \\ %& & ADVERSARIAL\_ATTACK\_DETECTED\\
        
        & type & InternalServerError \\
       
       & source &  String to indicate the source of the error. \\
       
       & data: attribute & String to indicate the attribute causing the error. \\
       
       & data: threshold & The threshold obtained from the configuration. \\

       & data: model\_confidence & The model confidence. \\

       & request-id & String to represent the unique identifier for the request. \\
       & error-id & String to represent the unique identifier for the error. \\
        \hline 
    \end{tabular}
    \caption{Structure of HTTP responses generated by the proposed protocol}
    \label{tab:http-responses}
    \vspace{-4mm}
\end{table*}
\subsection{Model-2-Software Communication Protocol}
The ML protocol employs HTTP status codes 200, 400 and 500~\cite{http_status_codes} to report successful outcomes and errors. An HTTP code 200 is returned when all necessary checks are satisfied, indicating a successful classification decision. This code signifies the successful completion of the process, and the response includes information such as the classification result, confidence level, and an explanation of the rationale behind the decision. An HTTP code 400 is returned if the input provided did not meet the necessary criteria or was incompatible with the system requirements. Also, an HTTP code 500 is returned if the process encounters failures, such as OOD data detection, or adversarial attack detection. This code indicates an internal server error and it comes with a comprehensive error message pinpointing the exact source of the problem. The message specifies the component where the error originated, offering software engineers a starting point for diagnostics. This approach enhances the transparency and interpretability of the ML model's decision-making process, providing valuable insights into both successful classifications and instances of error. The structure of these HTTP responses is illustrated in table~\ref{tab:http-responses}.

\subsection{ML Safeguards: Implementation Patterns}
Software systems could employ one or more ML models, which may have similar or distinct functions, often with complex interdependencies. A given ML model can be used by one or more software system. Safeguarding ML models is shared between two key stakeholders: the ML model provider, typically operating within an ML environment (Model-Safeguards), and the software engineers who work within the application development environment (Application-Safeguards).

Model-Safeguards place the responsibility on ML providers to develop safeguards designed to address the challenges associated with ML models use. It can be as either application-specific safeguards or common safeguard used across all applications. ML providers develop these guards based on their understanding of the training data, testing scenarios, model responses to unforeseen situations.

Application-Safeguards involve implementing safeguards on the application side, which gives flexibility to create safeguards for a specific ML model or cross-models safeguard that combines logic across all models (in particular when we use ensemble of ML models to achieve the task) . This pattern empowers application developers to adapt the safeguards that address the unique requirements of their applications and ML models they use, and this is especially valuable in cases where access to model details is limited.

Finally, both Model and Application Safeguards, which typically involves collaborative efforts between ML providers and application developers, working together to enhance system robustness using safeguarding techniques. Although this approach may involve increased costs, it is considered the optimal strategy for safeguarding ML models in production.

% details of the safeguards - techniques used 

\begin{figure*} %[htbp]
  \includegraphics[scale=0.5]{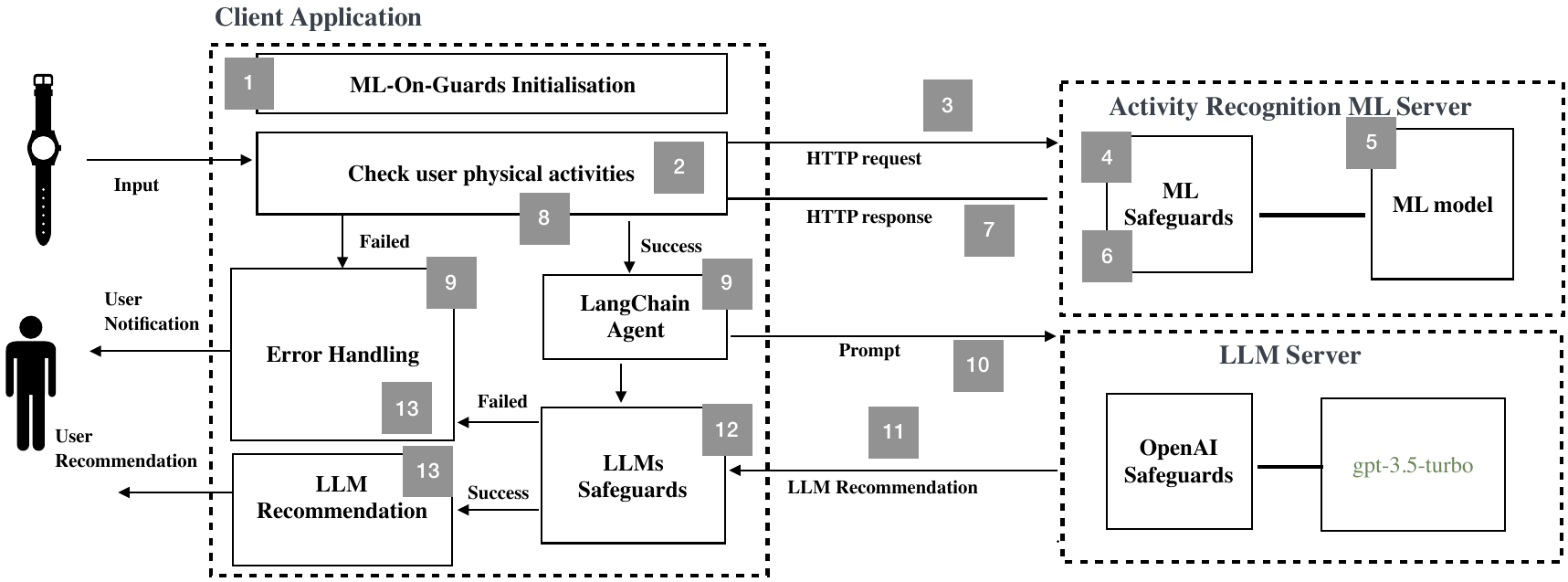}
  \caption{The flow of the MoveReminder application with the proposed ML-On-Rails protocol incorporated.}
   \label{fig:ml-protocol}
   \vspace{-4mm}
\end{figure*}

\section{Case study} 
\label{sec:study-design}

The MoveReminder is a mobile health application specifically designed to motivate individuals diagnosed with type-2 diabetes to decrease their sedentary lifestyle. Its central objective revolves around motivating users to adopt a more active lifestyle by offering personalized activity recommendations. This application utilizes a range of dynamic elements to create these recommendations. It considers variables like weather conditions, user specific location, current date and time, and user physical activity level during the day captured via a smart watch. By integrating these factors, the application ensures that the recommendation it provides is relevant and achievable for the user, ultimately helping them in their journey to reduce sedentary habits and improve their health. 

The MoveReminder application leverages a LangChain agent which has access to various specialized tools including; WeatherTool, LocationTool, DateTimeTool, ActivityRecognitionTool, LLMRecommendationGenerationTool, enabling it to orchestrate the process of fusing the tools to generate a personalised recommendation. There are two ML models in the application: an activity recognition model within the ActivityRecognitionTool and a GPT-3.5 LLM model within the LLMRecommendationGenerationTool. 

\noindent\textbf{Activity Recognition Model}
The activity recognition ML model takes raw data from the worn sensor (accelerometer readings), and classifies it into one of six distinct physical activities: downstairs, jogging, sitting, standing, upstairs, and walking.

\noindent\textbf{GPT-3.5 LLM model} 
The recommendation generation model is the GPT-3.5 LLM. We prompt the model to use the available information about the use state and environment to generate a recommendation as follows: \textit{Considering the user's current location (city), the time of day, the day of the week, the current weather conditions, and the recognition of the user's current activity, please provide a reasonable recommendation for an activity to help reduce the user's sedentary behavior.} In response, the LLM model formulates a personalized activity recommendation, as shown in Table~\ref{tab:example}.

% \begin{table}[!t]
%     \centering
%     \begin{tabular}{|l|}
%         \hline
%           Considering the user's current location (city), the time of day,\\ the day of the week, the current weather conditions, and the \\ recognition of the user's current activity, please provide \\ a reasonable recommendation for an activity to help reduce the \\
%           user's sedentary behavior. \\
%        \hline
%     \end{tabular}
%     \caption{The Recommendation Prompt} 
%     \label{tab:recommendation-prompt}
%     \vspace{-4mm}
% \end{table}

\begin{table}[!t]
    \centering
    \begin{tabular}{|l|}
        \hline
          Based on the location (Adelaide), time of the day (12:00 PM), day \\ of the week (Wednesday), weather (heavy rain, temperature 10°C,\\ humidity 40\%, wind speed 6.77 km/h), and the activity the user is \\ currently doing (jogging), I recommend the user to find an indoor \\ activity to reduce their sedentary behavior. Since it is early in the \\ morning and raining heavily, it may not be safe or comfortable to \\ continue jogging outdoors. Indoor activities such as yoga, \\ stretching, or home workouts can be great alternatives to \\ stay active while avoiding the rain. \\
       \hline
    \end{tabular}
    \caption{Example of the MoveReminder application output} 
    \label{tab:example}
     \vspace{-7mm}
\end{table}

% \begin{figure}
%   \includegraphics[trim={0 2.25mm 0 1.8mm},clip, width=\linewidth, scale=0.2]{images/moveReminder_output.png}
%   \caption{Example output of the MoveReminder application with ML-On-Rails. The equipped OOD detection safeguard identified the issue and triggered a server-side error labeled as "OUT\_OF\_DISTRIBUTION."}
%   \label{fig:app-output}
%   \vspace{-6mm}
% \end{figure}
\begin{figure}
  \includegraphics[width=\linewidth]{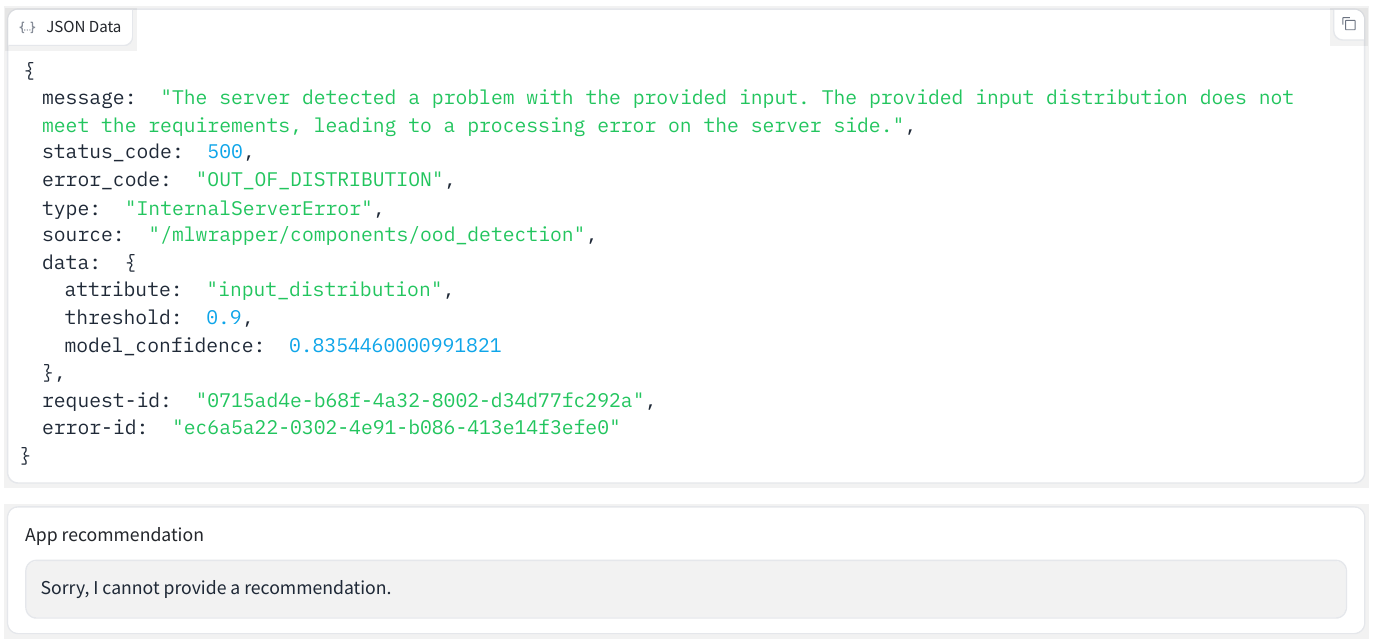}
  \caption{Example output of the MoveReminder application with ML-On-Rails. The equipped OOD detection safeguard identified the issue and triggered a server-side error labeled as "OUT\_OF\_DISTRIBUTION."}
  \label{fig:app-output}
  \vspace{-6mm}
\end{figure}

%\noindent\textbf{Importance of ML-On-Rails Protocol for the MoveReminder Application:} 
\noindent\textbf{MoveReminder Application with ML-On-Rails Protocol:} 

\noindent Figure~\ref{fig:ml-protocol} illustrates the flow of the MoveReminder application with the proposed protocol incorporated. There are three safeguard suites included; the activity recognition ML safeguards, OpenAI safeguards, and LLMs safeguards. The process initiates as the smartwatch sends data to the client application, which, in turn, forwards it to the activity recognition model server along with protocol configurations for physical activity classification. The ML server processes the input through specified safeguards based on developer-configured settings. Following this, an HTTP response is sent back to the client application. If the classification is successful, the LangChain agent utilizes relevant user activity information to prompt the LLM server for recommendations. After passing OpenAI safeguards, the client application receives and checks the recommendation for potential hallucinations using its own LLM safeguards. If all checks pass, the recommendation is delivered to the user. However, if any of the activity recognition ML safeguards fail, error handling notifies the user. So, In the event of failures, for example, input sensor data becomes corrupted, the activity recognition system may still attempt to classify the activity. In this scenario, the corrupted data is forwarded to the LangChain and the application would generate an irrelevant recommendation. With the ML-On-Rails protocol in place, the OOD detection safeguard is the most relevant for this kind of error after input validation. The corrupted data in the scenario has been identified by the OOD safeguard, indicating a deviation from training data. Consequently, the LLM will refrain from recommending any activities to the user as illustrated in Fig~\ref{fig:app-output}. This highlights the significance of the ML-On-Rails to ensure the reliability of the activity recommendations.

\section{Conclusion and future work}
\label{sec:conclusion}
The need for protocols safeguarding ML models is evident, serving as a framework to address evolving challenges in production. Continuous evolution and enhancement of the ML protocol, through the incorporation of additional safeguards, are crucial. In our future work, we plan a qualitative evaluation through a targeted survey involving software engineers and data scientists. The survey will directly capture insights and feedback from professionals, covering key elements of feasibility, effectiveness, and ease of use. Evaluation aspects include effectiveness of the ML protocol in addressing critical challenges, clarity of error/success messages, robustness enhancement, user-friendliness and seamless integration with existing systems.
\newpage

\bibliographystyle{ACM-Reference-Format}
\bibliography{references.bib}

\end{document}